\documentclass[aps,prd,twocolumn,superscriptaddress,showpacs]{revtex4}
\usepackage{graphicx}
\usepackage{dcolumn}
\usepackage{bm}

\newcommand{\apjs}{{Astrophys.~J.~Supp.}}

\newcommand{\mnras}{{Mon.~Not.~R.~Astron.~Soc.}}
\newcommand{\be}{\begin{equation}}
\newcommand{\ee}{\end{equation}}
\newcommand{\bea}{\begin{eqnarray}}
\newcommand{\eea}{\end{eqnarray}}

\begin{document}
\title{Constraining dark energy with cross-correlated
CMB and Large Scale Structure data}
\author{Pier-Stefano Corasaniti}
\email{pierste@phys.columbia.edu}

\affiliation{ISCAP, Columbia University,
550 West 120th Street,
New York, NY, 10027 (USA)}

\author{Tommaso Giannantonio}
\email{tommaso.giannantonio@roma1.infn.it}
\affiliation{Dipartimento di Fisica ``G. Marconi'', 
Universita' di Roma ``La Sapienza'', Ple Aldo Moro 5,
  00185, Roma, Italy.}

\author{Alessandro Melchiorri}
\email{alessandro.melchiorri@roma1.infn.it}
\affiliation{Dipartimento di Fisica ``G. Marconi'', 
Universita' di Roma ``La Sapienza'' and INFN, sezione di Roma, 
Ple Aldo Moro 5, 00185, Roma, Italy.}

\begin{abstract}
We investigate the possibility of constraining dark energy with 
the Integrated Sachs Wolfe effect recently detected by cross-correlating 
the WMAP maps with several Large Scale Structure surveys.
In agreement with previous works, we found that, under the assumption
of a flat universe, the ISW signal is a promising tool for
constraining dark energy.
Current available data put weak limits on a constant dark energy
equation of state $w$. We also find no constraints on 
the dark energy sound speed
$c_e^2$. For quintessence-like dark energy
($c_e^2=1$) we find $w<-0.53$, while
tighter bounds are possible only if
the dark energy is ``clustered'' ($c_e^2=0$), in such a case $-1.94<w<-0.63$
at $2\sigma$. 
Better measurements of the CMB-LSS correlation will be possible
with the next generation of deep redshift surveys. This will provide
independent constraints on the dark energy which are alternative to
those usually inferred from CMB and SN-Ia data. 
\end{abstract}
\maketitle

\section{Introduction}
\label{sec:intro}

The recent Wilkinson Microwave Anisotropy Probe (WMAP) satellite measurements
of the Cosmic Microwave Background (CMB) anisotropy spectra provide 
an accurate determination of several cosmological parameters \cite{Spergel}. 
In combination with complementary results from galaxy surveys \cite{2dF,SDSS},
these data strongly suggest that the present energy budget of the universe 
is dominated by an exotic form of matter which is also responsible 
for the present phase of acceleration as directly inferred from 
measurements of luminosity distance to Supernova type Ia \cite{PerlRiess}.

The presence of a cosmological constant term $\Lambda$ 
in Einstein's equation of General Relativity (GR) is the simplest
explanation for dark energy.
Remarkably the $\Lambda$ cold dark matter scenario ($\Lambda$CDM) 
is the minimal model to consistently account for all observations and
therefore has emerged as the standard model of cosmology. 

The WMAP temperature anisotropy maps have been also 
cross-correlated with several surveys of 
Large Scale Structure (LSS) distribution 
and a positive correlation signal has been detected
\cite{Gaztanaga03,Boughn0304,Scranton,Nolta04,Afshordi}.
This is indeed another success of the
$\Lambda$CDM model since the existence of
a positive correlation was 
already predicted nearly a decade ago by Crittenden and Turok 
as a test of flat $\Lambda$-cosmologies \cite{Crittenden}. 

Despite the simplicity of this concordance model, the nature of 
dark energy is far from being understood. In fact the existence of a small 
non-vanishing cosmological constant $\Lambda$ arises questions to which
present theoretical particle physics has found no consistent
explanation. On the other hand more exotic forms of matter such a scalar 
field cannot be {\em a priori} excluded. They have been proposed in several
context as candidate for dark energy
\cite{Wetterich,Ratra,Zlatev,Armendariz,Amendola}
and found to be consistent with current observations \cite{Quinteold}. 
Alternatively 
in a number of scenarios it has been suggested that what appears
as a dark energy component could be a manifestation
of a modified theory of gravity or a consequence of spatial extra 
dimensions \cite{Altern}. In
this paper we will consider only GR gravity with a dark energy fluid
whose energy momentum tensor violates the strong energy condition.

Within this framework the dark energy and the cosmological constant
differs for two main aspects: the latter behaves as a homogeneous fluid with 
a constant energy density, while the former is a non-homogeneous fluid
with a time dependent energy density and pressure. 
A simple way of describing these models is by specifying the equation of state
$w=p_{DE}/\rho_{DE}$, where $p_{DE}$ is the pressure and  
$\rho_{DE}$ is the energy density. 
The cosmological constant corresponds to the specific constant
value $w=-1$, while a general dark energy fluid may have a
time dependent equation of state $w(t)$. 
General covariance requires 
$\Lambda$ to be a homogeneous component, while dishomogeneities 
may occur in fluids with $w\neq -1$.
The clustering properties of different dark energy models
can be parameterized by an effective sound speed defined
as the ratio between the pressure to density perturbations in the rest
frame of dark energy, $c_e^2=\delta p_{DE}/\delta \rho_{DE}$ \cite{Hucs2}.  
Since we lack of a consistent theoretical formulation of
dark energy we are left with
constraining some phenomenological motivated form of $w(t)$ and $c_e^2$.

The cross-correlation between CMB and LSS offers a new complementary way of
constraining these parameters 
\cite{Cooray01,Afshordiman,HuScran,GarrLevon,Levon}. 
In fact the correlation between these data sets 
is consequence of the Integrated 
Sachs-Wolfe effect \cite{SW}, which has been shown to be 
a sensitive probe
of the evolution and clustering of dark energy 
\cite{CorasPRL,HutCoo}. 
It is therefore timely to investigate whether current
measurements of the CMB-LSS correlation
can already provide some novel constraints on the dark energy properties.

The paper is organized as follows : in Sec.~\ref{sec:theory} we introduce 
the ISW-correlation and study its dependence on dark energy. In
Sec.~\ref{sec:data} we describe the data analysis.
In Sec.~\ref{sec:results} we discuss the results and finally we draw
our conclusions in Sec.~\ref{sec:discussion}.

\section{Theory}
\label{sec:theory} 
In a flat dark energy dominated universe 
the gravitational potentials associated with the large scale structures
decay as consequence of the accelerated phase of expansion. CMB photons
which cross these regions acquire a shift which generates
temperature anisotropies. 
This is the so called Integrated Sachs-Wolfe (ISW) effect \cite{SW}.
A natural consequence of this mechanism is that if there is a clump of matter,
such as a cluster of galaxies in a given direction of the sky, 
we are likely to observe a spot in the 
corresponding direction of the CMB provided that
the CMB photons have crossed that region
during the accelerated epoch. We therefore expect to measure a positive
angular correlation between CMB temperature anisotropy maps and surveys 
of the large scale structures.

The ISW temperature fluctuation, $\Delta_{ISW}$, 
in the direction $\hat{\gamma_1}$ is
given by:
\be
\Delta_{ISW}(\hat{\gamma_1})=-2 \int e^{-\tau(z)}\frac{d\Phi}{dz}(\hat{\gamma_1},z) dz,
\label{isw}
\ee
where $\Phi$ is the Newtonian gauge gravitational potential and 
$e^{-\tau(z)}$ is the visibility function
to account for a possible suppression due to early reionization.

The density contrast corresponding to a clump of luminous matter observed
by a given survey in the direction $\hat{\gamma_2}$ is:
\be
\delta_{LSS}(\hat{\gamma_2})=b \int \phi(z) \delta_m(\hat{\gamma_2},z) dz,
\ee
with $\delta_m$ the matter density perturbation, $b$ the galaxy bias and 
$\phi(z)$ is the selection function of the survey.

The 2-point angular cross-correlation is defined as
\be
C^X(\theta)=\langle\Delta_{ISW}(\hat{\gamma_1})\delta_{LSS}(\hat{\gamma_2})\rangle,
\label{cross}
\ee
where the angular brackets denote the average over the ensemble and
$\theta=\vert \hat{\gamma_1}-\hat{\gamma_2}\vert$.
For computational purposes it is convenient to decompose 
$C^X(\theta)$ into the Legendre series such that,
\be
C^X(\theta)=\sum_{l=2}^{\infty}\frac{2l+1}{4\pi}C_l^{X}P_l(\cos(\theta),
\label{cxpl}
\ee
where $P_l(\cos\theta)$ are the Legendre polynomials and $C_l^{X}$ 
is the cross-correlation power spectrum given by \cite{GarrLevon,Levon}:
\be
C_l^X=4\pi\frac{9}{25}\int \frac{dk}{k}\Delta_{\mathcal{R}}^2 I^{ISW}_l(k)I^{LSS}_l(k),
\ee 
where $\Delta_{\mathcal{R}}^2$ is the primordial power spectrum.
The integrand functions $I^{ISW}_l(k)$ and $I^{LSS}_l(k)$ are defined 
respectively as:
\begin{eqnarray}
I^{ISW}_l(k)&=&-2\int e^{-\tau(z)} \frac{d\Phi_k}{dz} j_l[k r(z)] dz\\
I^{LSS}_l(k)&=&b\int  \phi(z) \delta^k_m(z)j_l[k r(z)] dz,\label{growth}
\end{eqnarray}
where $\Phi_k$ and $\delta_m^k$ are the Fourier components 
of the gravitational potential and matter perturbation respectively, 
$j_l[k r(z)]$ are the spherical Bessel functions and $r(z)$ is the 
comoving distance at redshift $z$.
 
Direct measurements of the cross-power spectrum $C_l^X$ 
are more robust for likelihood parameter estimation since these data would 
be less correlated than measurements of $C^X(\theta)$.
However current observations provide only estimates of the 
angular cross-correlation function, for this reason we focus 
on $C^X(\theta)$.

In order to better illustrate the model dependence of the ISW we
compute the angular cross-correlation function for different values of $w$
and $c_e^2$.  We first compute the cross-power spectrum of a given
model using a properly modified version of
CMBFAST code \cite{seljak}, then we evaluate $C^X(\theta)$ using
Eq.~(\ref{cxpl}). In this way the monopole and dipole contribution
to the angular correlation function are subtracted by construction. 
We have implemented the
dark energy perturbation equations as described in \cite{Bean}.
In what follows we assume a flat universe
with Hubble parameter $h=0.68$, $\Omega_{DE}=0.7$, baryon density
$\Omega_b h^2=0.024$, scalar spectral index $n_s=1$, no reionization. 
The amplitude of the primordial fluctuations $A_s=0.86$ 
(as defined in \cite{verde}) and the bias $b=1$.
We consider a Gaussian selection function peaked at
$z=0.1$ with variance $\sigma_{\phi}=0.15$.

\begin{figure}[t]
\includegraphics[scale=0.4]{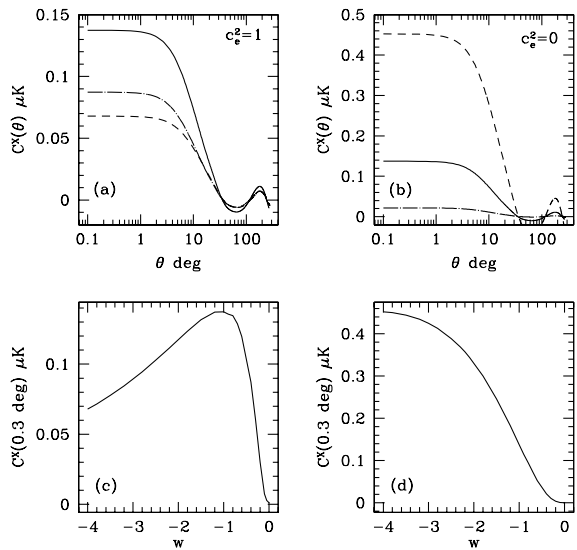}
\caption{Angular cross-correlation function for $c_e^2=1$ (a) and 
$c_e^2=0$ (b), the different lines corresponds to $w=-0.8$ (solid), 
$w=-0.4$ (long dash-dot) and $w=-4$ (short dash). Amplitude 
of the angular cross-correlation at the plateau as function of $w$
for $c_e^2=1$ (c) and $c_e^2=0$ (d).
}
\label{crossz0.1}
\end{figure}

In Figure~\ref{crossz0.1}a we plot the angular cross-correlation
for the case $c_e^2=1$ and in Figure~\ref{crossz0.1}b for $c_e^2=0$.
The different lines correspond to $w=-0.8$ (solid), $w=-0.4$ (long dash-dot) 
and $w=-4$ (short dash).

We notice that at small angles the angular
cross-correlation is characterized by a nearly constant plateau,
while it rapidly vanishes at larger angles ($\theta>10^\circ$).
The overall amplitude of the signal up to angles $\theta\sim few^{\circ}$ 
is particularly sensitive to both $w$ and $c_e^2$. 
This can be better seen in Figure~\ref{crossz0.1}c 
and Figure~\ref{crossz0.1}d where we plot
the amplitude of cross-correlation at 
the plateau, $C^X(0.3^\circ)$, as function of $w$ 
in the case $c_e^2=1$ and $c_e^2=0$ 
respectively. 

For $c_e^2=1$, the amplitude has a maximum around $w=-1$
and slowly decreases for decreasing values of $w$, 
while it rapidly falls to zero for $w\rightarrow 0$ (Fig.~\ref{crossz0.1}c).
This is because the dark energy contribution to the ISW effect 
is mainly due to the background expansion. 
In fact for models with $w>-1$, as
$w\rightarrow 0$ the dark energy driven expansion is less
accelerated and tends
to the matter dominated behavior. Hence the variation of the gravitational
potentials is smaller and consequently produces a
negligible amount of ISW as $w\rightarrow 0$.
Similarly for models with $w<-1$, the dark energy affects 
the expansion later than in models with $w\ge-1$. 
This effectively extends the period
of matter domination which leads to a lower ISW signal.

Since a smaller ISW signal can be compensated by increasing the amount
of dark energy density $\Omega_{DE}$, we expect a precise degeneracy line
in the $\Omega_{DE}-w$ plane. In particular lower negative values of $w$ 
will be counterbalanced by higher values of $\Omega_{DE}$.

On the contrary for $c_e^2=0$, the amplitude of the cross-correlation 
is a monotonic decreasing function of $w$ (Fig.~\ref{crossz0.1}d). 
In this case the decay of the gravitational potential is sensitive to the 
clustering of dark energy which is more effective as $w$ decreases.
Thus the amplitude
of the ISW increases as $w$ decreases 
\cite{Bean,JochenLewis}. We therefore expect the degeneracy 
in the $\Omega_{DE}-w$ plane to be orthogonal to the previous
case. In fact increasing $\Omega_{DE}$ will compensate for larger
values of $w$. 

This trend hold independently of the selection function as long as 
it is centered in a range of redshifts up $z\sim0.7-0.8$ 
for models with $w\ge-1$.
However one might expect this to not be the general situation 
in the case of dark energy models with a time dependent equation of state 
\cite{Levon}. 

\begin{figure}[t]
\includegraphics[scale=0.4]{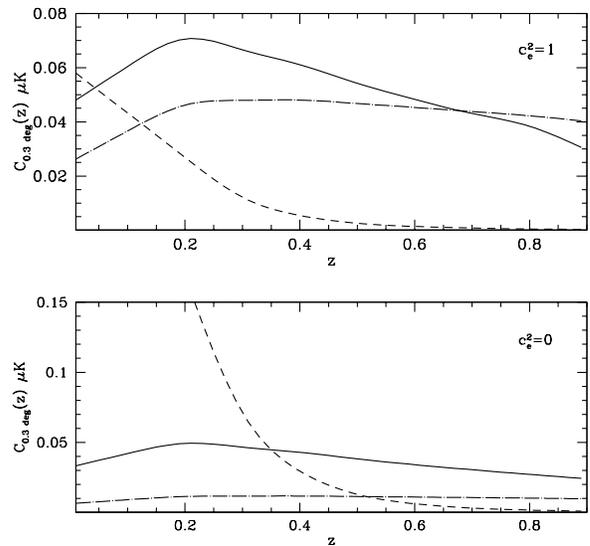}
\caption{Amplitude of the angular cross-correlation at the plateau as function
of the redshift. A Gaussian selection function with width 
$\sigma_{\phi}=0.05$ has been used. 
The different lines correspond to models as in 
Figure~\ref{crossz0.1}.
}
\label{ctw}
\end{figure}

In Figure~\ref{ctw} we plot the amplitude of the angular cross-correlation 
at the plateau as function of the redshift. 
We have used a Gaussian function centered at different redshifts in
steps of $0.1$ and with a width $\sigma_{\phi}=0.05$. 
The upper (lower) panel shows the case $c_e^2=1$ ($c_e^2=0$), the different
lines correspond to the same models as in Figure~\ref{crossz0.1}.
It can be seen that for $z\gtrsim 0.2$ the signal decreases 
with the redshift in way that is strongly dependent on the dark energy
parameters. Therefore redshift measurements of the cross-correlation
are a potentially powerful tool to distinguish between different dark
energy models.

We may also notice that having used a narrower selection function
the amplitude of the signal is systematically 
smaller than that found in previous cases. 
For instance for $w=-0.8$ and $c_e^2=1$ we have $C(0.3^{\circ})=0.06$$\mu{K}$
at $z=0.1$ (Fig.\ref{ctw} upper panel),
which is smaller than $C(0.3^{\circ})=0.14$$\mu{K}$ 
(see Fig.\ref{crossz0.1}a) obtained using 
a wider selection function ($\sigma_{\phi}=0.15$).

Hence a sharper selection function gives a smaller 
cross-correlation signal, eventually leading to larger uncertainties. 
On the other hand increasing the number of uncorrelated redshift bins 
would allow a better reconstruction of 
the redshift evolution of the cross-correlation.
 
This suggests that there could be an optimal way
of designing a large scale structure survey which 
maximizes the discrimination power of the cross-correlation independently
of the model of dark energy. This can be achieved using Integrated Parameter
Space Optimization (IPSO) techniques \cite{BruceIPSO}. We leave
this interesting possibility to further investigation and
we refer to \cite{Levonetal} for a detailed analysis of  
dark energy parameter forecast from future cross-correlation data.

\section{Method and Data}
\label{sec:data}
We perform a likelihood analysis using the collection 
of data presented in Gaztanaga {\em et al.} \cite{gazt}, 
which has the advantage of being publicly available and easy to implement.
These data consist of measurements of the average angular cross-correlation
around $\theta=5^{\circ}$ between WMAP temperature anisotropy maps
 and several LSS surveys. 
The angular average around $\theta=5^{\circ}$
ensures that the signal is not 
contaminated by foregrounds such as the
SZ or lensing effects which are relevant at smaller angles 
($\theta\lesssim 1^{\circ}$). Taking the average 
does not wash out the sensitivity
on the dark energy parameter since, as 
we have shown in Section~\ref{sec:theory}, the amplitude
of the cross-correlation around $\theta=5^{\circ}$ is still strongly
dependent on the value of $w$ and $c_e^2$.
 
Possible systematic contaminants, 
such as extinction effects, seem not to affect these data and 
for a more detailed discussion we refer to \cite{gazt}.

The data span a range of redshift $0.1<z\lesssim 1$ and
for each redshift bin the data include an estimate of the galaxy bias
with $20\%$ errors. These biases are inferred by comparing the 
galaxy-galaxy correlation function of each experiments 
with the expectation of LCDM best fit model to WMAP power spectra.

The amplitude of the scalar density perturbation $A_s$ is an overall
normalization factor which we can marginalize over, 
on the contrary prior knowledge of the bias is required. 
In fact a scale and/or redshift dependent bias can in principle mimic
the redshift evolution of the cross-correlation predicted by different
dark energy models. 
It is therefore necessary to
have an independent estimate of $b$, for instance by
combining weak lensing information \cite{HuJain} or measurements
of the matter power spectrum with
the bispectrum \cite{Dolney}. 

One of the advantages of testing dark energy with the cross-correlation is 
that it is insensitive to other parameters which limits common
dark energy parameter extraction analyses involving 
CMB temperature and polarization anisotropy
spectra \cite{Coras04}. For instance the ISW correlation is not affected 
by a late reionization or by an extra background of
relativistic particles which change the CMB spectra through the early-ISW
(see e.g. \cite{bowen}). The ISW-correlation is also 
independent of the amplitude of tensor modes and
depends uniquely on the scalar perturbations, since
a primordial background of gravity waves is uncorrelated
with present large scale structure distribution.

There is little sensitivity to the scalar spectral index $n_s$, while
the dependence on the baryon density $\Omega_b$ can be non-negligible. 
In fact the presence of baryons inhibits the growth of CDM fluctuations
between matter-radiation equality and photon-baryon decoupling \cite{Hu}
causing the matter power spectrum to be suppressed on scales $k>k_{eq}$
for increasing values of $\Omega_b$
($k_{eq}$ is the scale which enters the horizon at the equality). 
Over the range of scales which
contribute to the ISW-correlation ($k\sim 0.01$) 
the sensitivity on $\Omega_b$ is still present.
In order to limit the number of likelihood parameters
we therefore assume a Gaussian prior on the baryon density 
$\Omega_b h^2=0.0216\pm 0.002$
consistent with WMAP and Big-Bang Nucleosynthesis bounds
and consider the following set parameters:
the matter density $\Omega_m$, the Hubble constant $h$, the equation
of state $w$ and the sound speed of dark energy $c_e^2$. 
We assume a scale invariant primordial spectrum $n_s=1$ and fix
the optical depth to WMAP best fit value
$\tau=0.17$ (again the ISW is not particularly affected by a change
in those parameters). We marginalize over the normalization amplitude
$A_s$, although we found no difference assuming the WMAP value. 
In fact changing $A_s$ shifts the overall amplitude of the angular 
cross-correlation of the same amount over different redshifts but it 
does not change the redshift dependence of the signal.

Since the experimental data are corrected for the bias 
by comparing the measured galaxy-galaxy correlation function
in each redshift bin to the WMAP best fit model, 
we rescale these biases to each of the dark energy model
in our database as described in \cite{gazt}.

We compute for each theoretical model
the angular cross-correlation as described in Sec.~\ref{sec:theory}
using the selection function 
\begin{equation}
\phi^i(z)\sim z^2 \exp{[-(z/\bar{z}_i)^{1.5}]},\label{sel}
\end{equation} 
where $\bar{z}_i$ is the median redshift of the $i$-th survey.
Then following \cite{gazt}, we compute the average cross-correlation
in the $i$-th bin, $\bar{C}_i^X$,
around $\theta\sim 5^{\circ}$.
 
The data points are an average over
angles and are inferred from surveys whose selection functions
may overlap in redshift space, hence they are not independent
measurements and indeed are affected by a certain degree of correlation.
Since we have no access to the raw data 
we have no way of accounting for the first type of correlation, while
using Eq.~(\ref{sel}) we can estimate the correlation between
different redshift bins. We compute the correlation matrix 
$\rho=\{\rho_{ij}\}$, where $\rho_{ij}$ is the fraction 
of overlapping volume between the $i$-th and $j$-th surveys (i.e. the diagonal
components are $\rho_{ii}=1$). 
To be more conservative we have assumed that the different surveys 
cover the same fraction of sky, in general this is not the case
and the fraction of overlapping volume can be smaller.
We found that only two data points are highly correlated,
since their selection functions overlap for about $70\%$ in redshift space, 
while the correlation among the remaining data points are less than $20\%$.

We compute a likelihood function $\cal L$ 
defined as

\begin{equation}
-2\log{\cal L}=\chi^2=\sum_{ij}(\bar{C}_i^X-{\hat C}_i^X)
M_{ij}^{-1}(\bar{C}_j^X-{\hat C}_j^X),
\end{equation}
where ${\hat C}^X_i$ are the data and
$M_{ij}^{-1}=\rho_{ij}/(\sigma_i \sigma_j)$ is our estimate of 
the inverse of the covariance matrix, with $\sigma_i$ the measured 
uncertainty in the $i$-th bin. 

We also use the SN-Ia ``Gold'' data \cite{riess} to derive complementary
constraints on the dark energy parameters and compare with those derived
from the ISW correlation.

\section{Results}
\label{sec:results}

\begin{figure}[t]
\includegraphics[scale=0.35]{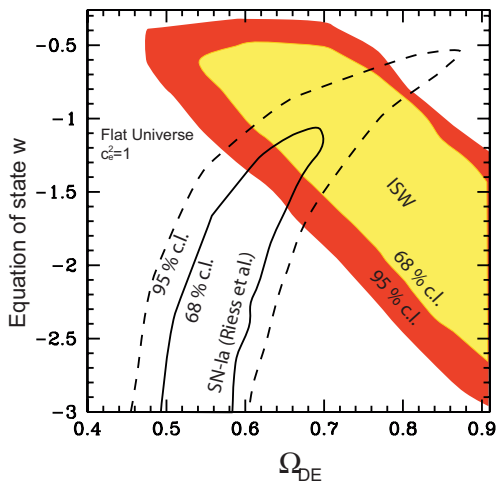}
\caption{Two-dimensional marginalized likelihoods 
on $\Omega_{DE}-w$. The yellow and red area correspond to
$1$ and $2\sigma$ limits inferred from the ISW data for $c_e^2=1$.
Solid and dash lines represent the $1$ and $2\sigma$ contours from 
the SN-Ia data.
}
\label{olwcs1}
\end{figure}

\begin{figure}[h]
\includegraphics[scale=0.35]{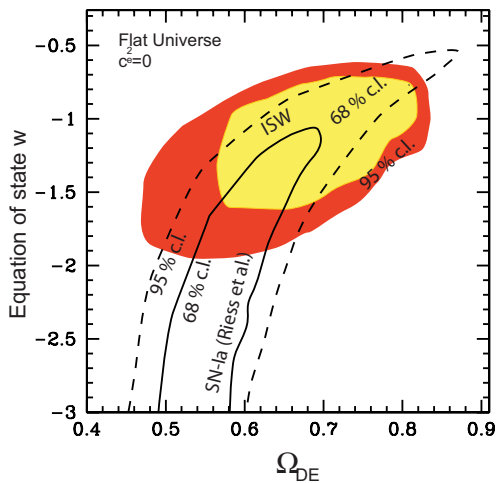}
\caption{As in Figure~\ref{olwcs1} with $c_e^2=0$ prior.
}
\label{olwcs0}
\end{figure}

We now discuss the results of the likelihood analysis and show
the marginalized constraints on the dark energy parameters. As expected
we found that the results depend on the dark energy
clustering. For instance in Figure~\ref{olwcs1} 
and \ref{olwcs0} we plot the two-dimensional marginalized
likelihoods in the $\Omega_{DE}-w$ plane for $c_e^2=1$ and $c_e^2=0$
respectively. The yellow and red contours correspond to $1$ and $2\sigma$
limits. We also plot the marginalized likelihood inferred from the SN-Ia 
data.

It appears evident that the derived limits on $w$ strongly depend on
the prior value of $c_e^2$. In particular varying $c_e^2$
changes the direction of the degeneracy between $\Omega_{DE}$ and $w$.
This is a natural consequence of the discussion presented in 
Section~\ref{sec:theory}. Namely, for $c_e^2=1$ decreasing $w$ 
causes a lower ISW signal which can be compensated by larger values
of $\Omega_{DE}$, while the opposite occurs for $c_e^2=0$.
Therefore for $c_e^2=1$ the ISW data provide only
a weak upper limit on the equation of state, $w <-0.53$ at $68\%$ confidence
level \footnote{The $1\sigma$ upper limit we found on $w$ is much 
weaker than that obtained in \cite{Vielva}, where the 
WMAP data have been correlated with the NVSS in the wavelet space.}. 
On the contrary assuming $c_e^2=0$ we obtain a tighter constraint 
$-1.94< w<-0.63$ at $2\sigma$. 

We can also notice that for $c_e^2=0$ the combination
of the ISW likelihood with the SN-Ia one has little effect 
in determining $w$, since the $\Omega_{DE}-w$ degeneracy of 
the luminosity distance lies in the same 
direction of the ISW cross-correlation. This is not the case for $c_e^2=1$,
where the degeneracy lines are orthogonal.
Therefore the combined
likelihoods provide a more stringent bound, 
$-1.72 < w < -0.53$ at $95 \%$. 

These limits are stable 
assuming a standard prior on the matter
density. For instance for $\Omega_m=0.27\pm0.04$
we obtain $-1.51<w<-0.72$ for $c_e^2=0$ and
$-1.81<w<-0.53$ for $c_e^2=1$ at $95 \%$.

The constraints we have inferred so far would be $~15-20\%$
tighter if we had ignored the correlations amongst
different redshift bins. In fact, as we have discussed in the previous
Section, the level of correlation of current data is still non-negligible.
On the contrary the next generation of surveys will be 
characterized by more localized
selection functions and provide uncorrelated cross-correlation 
measurements.

\begin{figure}[th]
\includegraphics[scale=0.35]{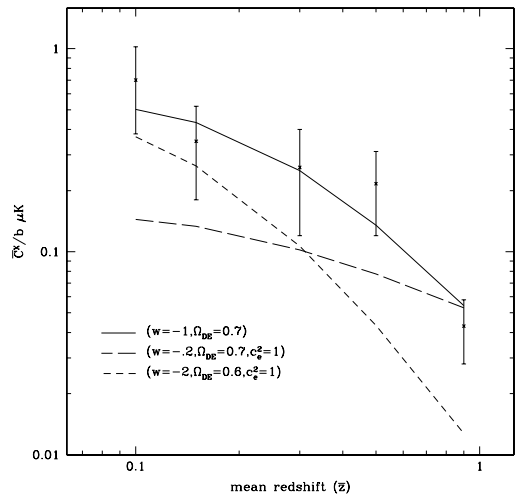}
\caption{Dots with errorbars are the different measurements of $\bar{C}^X/b$,
the solid line corresponds to the LCDM best fit model, the long-dash and
short-dash lines show the case ($w=-0.2,\Omega_{DE}=0.7$) and 
($w=-2,\Omega_{DE}=0.6$) respectively.} 
\label{fits}
\end{figure}

In Figure~\ref{fits} we plot the $5$ data points and 
three different model predictions of 
the average angular cross-correlation normalized to the bias and
the data. The solid line corresponds to the LCDM best fit model,
we also plot a non-accelerating dark energy dominated model (long-dash line)
and a phantom model (short-dash) which are disfavored by the data.

\begin{figure}[th]
\includegraphics[scale=0.35]{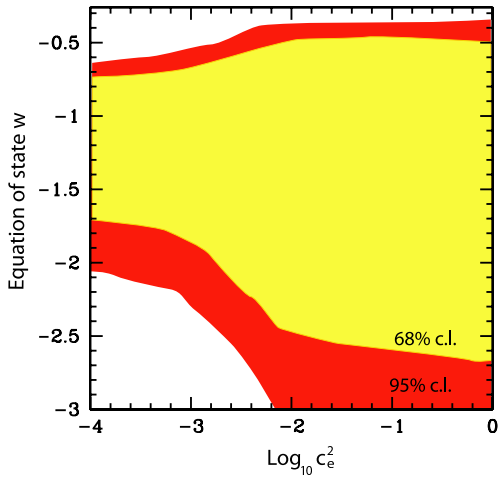}
\caption{Two-dimensional marginalized likelihood 
on $\log_{10}c_e^2-w$ from ISW-correlation data. The yellow and red
contours correspond to $1$ and $2\sigma$ limits.}
\label{wcs2}
\end{figure}

In Figure~\ref{wcs2} we plot  
the marginalized $1$ and $2\sigma$ contours
in the $\log_{10}{c_e^2}-w$ plane, as it can be seen the dark 
energy sound speed remains unconstrained. This is consistent with
the results from the CMB data analysis by Weller and 
Lewis \cite{JochenLewis} and
only at $2\sigma$ with Bean and Dore \cite{Bean}.

\begin{figure}[th]
\includegraphics[scale=0.35]{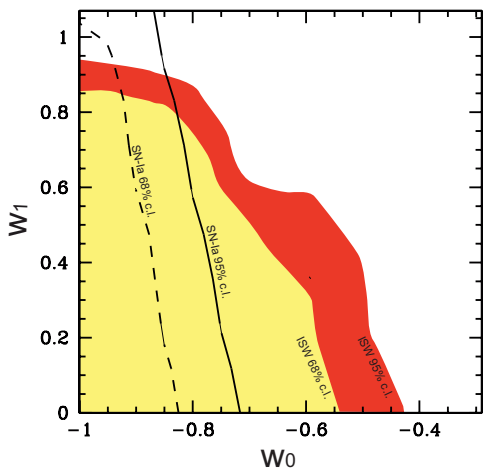}
\caption{$1$ and $2\sigma$ limits from ISW-correlation 
(yellow and red contours) and SN-Ia 
luminosity distance (dash and solid lines) on $w_0-w_1$.}
\label{wdynwa}
\end{figure}

We have also extended our analysis to constrain a class of
slowly varying dark energy models described by a parameterization of 
the equation of state which linear in the scale factor: $w(a)=w_0+(1-a)w_1$
\cite{PolarskiLind}.
 
This part should be considered as a simple exercise 
since current cross-correlation data are not accurate enough to allow
us to constrain more than two parameters. We therefore 
assume an $\Omega_m=0.3$ prior and we also restrict 
our analysis to models satisfying the Weak Energy 
Condition (WEC), $w_0>-1$ and $0 < w_1 <1$ such that we can
consistently account for the dark energy perturbations.

In Figure~\ref{wdynwa} we plot the marginalized $1$ and $2\sigma$ contours
in $w_0-w_1$ plane. We find that the cross-correlation data 
provide only weak upper limits on these parameters, 
$w_0<-0.53$ and $w_1<0.84$.
If we limit our analysis to models with $c_e^2=0$ 
the constraints are $w_0<-0.82$ and $w_1<0.84$ at $68 \%$.
These limits are in agreement with similar constraints
from SN-Ia as derived in previous analysis 
(see for instance \cite{BruceCoraKunz}).

\section{Discussion}
\label{sec:discussion}

The cross-correlation between CMB and LSS data provides a new complementary
way of testing dark energy. By isolating the ISW contribution to the
CMB anisotropies, the correlation is a sensitive probe of the dark energy
properties. In this paper we have studied its dependence on the dark
energy equation of state $w$ and sound speed $c_e^2$.
In particular we have shown that the redshift dependence of the 
cross-correlation signal may discriminate between different dark 
energy models and provide
constraints on the dark energy parameters alternative to those inferred from
cosmological distance measurements. In addition the ISW correlation is
insensitive to a number of cosmological parameters which usually
limits the dark energy parameter estimation from CMB data alone.
A precise knowledge of the galaxy bias is necessary for this 
type of measurements to be competitive. In fact a redshift or scale
dependent bias can in principle mimic the effect induced by dark
energy. 

We have also briefly reviewed the current observational status and
inferred constraints on $w$ and 
$c_e^2$ using the current limited datasets. We found that, even under 
several theoretical and optimistically experimental assumptions, the
actual constraints are weak. However, the presence of
a dark energy component is clearly significant in the data and
interesting. The constraints on the equation of state strongly depends
on the dark energy sound speed. In the case $c_e^2=1$ the ISW data provide
only a weak upper limit on $w<-0.53$ at $1\sigma$, while tighter bounds
are obtained assuming $c_e^2=0$. In agreement with previous independent works
based on CMB data alone we found no constraints on the dark energy
sound speed. Slowly varying dark energy models are also consistent with
current ISW data but not significantly preferred.
The upcoming deep redshift surveys such as LSST, KAOS or ALPACA are 
optimal datasets for studying the redshift evolution
of the CMB-LSS correlation and provide alternative
constraints on the dark energy parameters.

\acknowledgements
We are particularly grateful to Levon Pogosian for the very
helpful suggestions and for checking our 
numerical results with his ISW-correlation code.
We would like to thank also Enrique Gaztanaga and Marilena Loverde for
comments and discussions. P.S.C. is supported by Columbia Academic Quality
Fund, A.M. is supported by MURST through COFIN contract no. 2004027755.

\end{document}